\documentclass[10pt,superscriptaddress,twocolumn,amsmath,amssymb,aps,prl,showpacs]{revtex4-1}
\usepackage{mathrsfs}
\usepackage{graphicx}
\usepackage{dcolumn}
\usepackage{bm}
\usepackage{amssymb}
\usepackage{amsmath}
\usepackage{paralist}

\newcommand{\tmop}[1]{\ensuremath{\operatorname{#1}}}

\def\be{\begin{equation}}
\def\ee{\end{equation}}
\def\bea{\begin{eqnarray}}
\def\eea{\end{eqnarray}}

\begin{document}

\title{ Quantum criticality     of spinons
}

\author{Feng He}
\affiliation{State Key Laboratory of Magnetic Resonance and Atomic and Molecular Physics,
Wuhan Institute of Physics and Mathematics, Chinese Academy of Sciences, Wuhan 430071, China}
\affiliation{University of Chinese Academy of Sciences, Beijing 100049, China.}

\author{Yu-Zhu Jiang}
\affiliation{State Key Laboratory of Magnetic Resonance and Atomic and Molecular Physics,
Wuhan Institute of Physics and Mathematics, Chinese Academy of Sciences, Wuhan 430071, China}

\author{Yi-Cong Yu}
\affiliation{State Key Laboratory of Magnetic Resonance and Atomic and Molecular Physics,
Wuhan Institute of Physics and Mathematics, Chinese Academy of Sciences, Wuhan 430071, China}
\affiliation{University of Chinese Academy of Sciences, Beijing 100049, China.}

\author{H.-Q. Lin}
\email[e-mail:]{haiqing0@csrc.ac.cn}
\affiliation{Beijing Computational Science Research Center, Beijing 100193, China}


\author{Xi-Wen Guan}
\email[]{xiwen.guan@anu.edu.au}
\affiliation{State Key Laboratory of Magnetic Resonance and Atomic and Molecular Physics,
Wuhan Institute of Physics and Mathematics, Chinese Academy of Sciences, Wuhan 430071, China}
\affiliation{Center for Cold Atom Physics, Chinese Academy of Sciences, Wuhan 430071, China}
\affiliation{Department of Theoretical Physics, Research School of Physics and Engineering,
Australian National University, Canberra ACT 0200, Australia}

\date{\today}


\begin{abstract}

The free fermion nature of interacting spins  in one dimensional (1D) spin chains  still lacks a rigorous study.
In this letter  we show  that the length-$1$ spin strings significantly  dominate   critical  properties  of  spinons, magnons and  free fermions  in the 1D antiferromagnetic spin-1/2  chain.
Using the  Bethe ansatz solution  we  analytically  calculate exact scaling functions of thermal and magnetic  properties of the model,  providing  a rigorous  understanding  of  the quantum criticality of spinons.
It turns out   that   the double peaks  in  specific heat elegantly mark two crossover temperatures fanning out from the critical point, indicating three quantum phases:  the Tomonaga-Luttinger liquid  (TLL), quantum critical  and fully polarized ferromagnetic phases.
For the TLL phase,   the  Wilson ratio $R_W=4K_s$ remains 
almost temperature-independent, here $K_s$ is the Luttinger parameter.
Furthermore, applying our results we precisely determine the  quantum scalings and critical exponents of all magnetic properties in the 
ideal  1D spin-1/2 antiferromagnet Cu(C${}_4$H${}_4$N${}_2$)(NO${}_3$)${}_2$  recently studied in Phys. Rev. Lett. {\bf 114}, 037202 (2015)].
We  further find that the magnetization peak used in experiments 
is not a good quantity to map out the  finite temperature TLL phase boundary.

\end{abstract}

\maketitle

 %
 %
 %

\begin{figure}[t]
\begin{center}
  \includegraphics[width=0.92\linewidth]{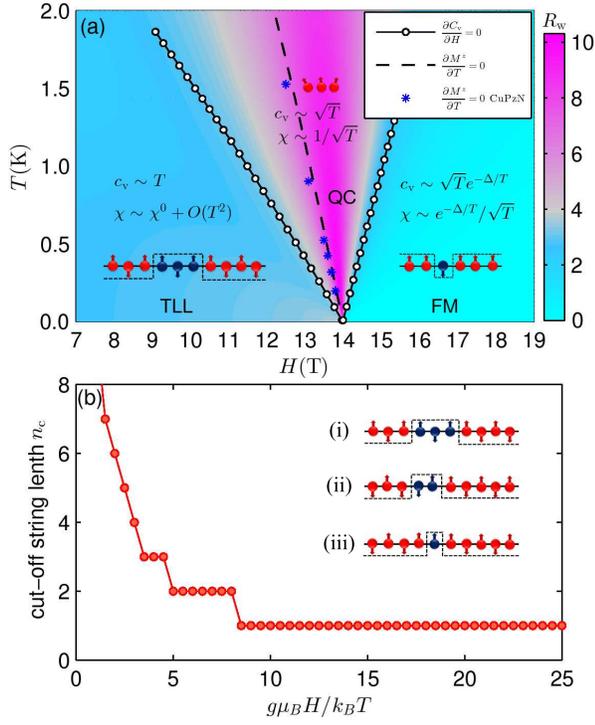}
  \end{center}
  \caption{
  (a) Contour plot of the Wilson ration 
  $R_\mathrm{W}$ in the $T-H$ plane. Without losing generality we used the realistic  coupling constant $2J=10.81K$ and the 
  Lande factor $g=2.3$ of  the spin-1/2 compound CuPzN.   It   maps out  quantum scalings of the TLL, the quantum critical (QC) region and the fully polarized ferromagnetic (FM) phase. The dotted solid lines fanning out from the saturation field  $H_s=4J$
  show the peak positions of the specific heat.
  The  black dashed line shows   the magnetization peaks determined from the TBA equations (\ref{TBA}). The blue stars show the experimental magnetization peaks.
  (b) The cut-off  string length $n_c$ versus
  the  energy scale $g\mu_B H/(k_BT)$ at  an accuracy of the order of $10^{-6}$. The cut-off $n_c$  shows stir-like features
  at low temperatures. The inset shows 
  three schematic spin configurations:  (i) $M^z=1$ and $2$ spinons;   (ii) $M^z=0$, $\nu_2=1$ and  $2$ spinons; (iii) $M^z=1$, $\nu_2=1$ and  $4$ spinons, see \cite{Supp}.}
  \label{Fig1}
\end{figure}

Of central importance to the study of the 1D spin-1/2 antiferromagnetic Heisenberg chain is  the  understanding of spin excitations  \cite{Yang:1966a,Faddeev:1981,Haldane:1981,Affleck:1986a,Takahashi:1999,WangYP:2015,Johnston:2000,Tennant:1995,Lake:2005,Mourigal:435,Lake:2013,Zheludev:2008,Stone:2003}.
Elementary spin excitations in this 
model  may exhibit quasi-particle behaviour which is  described by  spinons carrying half a unit of spin. 
Such fractional quasiparticles  are responsible for 
the TLL   in  the model  \cite{Affleck:1986,Cardy:1986,Giamarchi:2004}.

Regarding to the Bethe ansatz solution of   the 1D spin-1/2 chain, a  significant development is Takahashi's  discovery of  spin string patterns \cite{Takahashi:1971}, i.e., magnon  bound states with different string lengths.
Takahashi's  spin strings give one
full  access to the thermodynamics of the model  through  Yang and Yang's grand canonical approach \cite{Yang:1969},  namely
 the so-called thermodynamic  Bethe ansatz (TBA) equations \cite{Takahashi:1971}.
%
%
However, the problems of how  such  spin strings  determine the free fermion nature of  spinons 
and how
spin strings comprise universal scalings  of thermal and magnetic properties 
still lack 
a rigorous  understanding. In this paper we present a full answer to these questions.


Using spin string solutions to the TBA equations, we obtain 
the following results: I) we obtain exact scaling functions, critical exponents and a benchmark of quantum magnetism for  the 1D spin-1/2 Heisenberg chain, revealing the 
microscopic origin of the quasiparticle spinons, 
free fermions and magnons that emerge 
in different physical regimes;
II) We find that the Wilson ratio \cite{Som28,Wil75}, the ratio  between the susceptibility $\chi$ and the specific heat $c_v$ divided by the
temperature $T$,
$R_\mathrm{W}=\frac{4}{3}\left(\frac{\pi k_B}{g \mu_B
}\right)^2\chi/(c_v/T)$, 
significantly characterises   the TLL of spinons  and  marks the crossover  temperature between the quantum critical phase and the TLL  \cite{Kono:2015}, see  Fig.~\ref{Fig1}.
When
the magnetic field is larger than the saturation field, dilute magnon behaviour is evidenced by the 
exponential decay   of the susceptibility;
III) Using our  analytical and numerical results we  precisely determine the  quantum scalings and magnetic properties  of  the  ideal spin-1/2 antiferromagnet   Cu(C${}_4$H${}_4$N${}_2$)(NO${}_3$)${}_2$ (denoted by 
CuPzN for short)  \cite{Kono:2015}. We also find that the magnetization peak used  in experiment \cite{Kono:2015,Ruegg:2008,Shaginyan:2016} is not a good quantity to map out the finite temperature TLL phase boundary. Instead one should use  the Wilson ratio or the specific heat peaks.
\begin{figure}[t]
\begin{center}
  \includegraphics[width=0.9\linewidth]{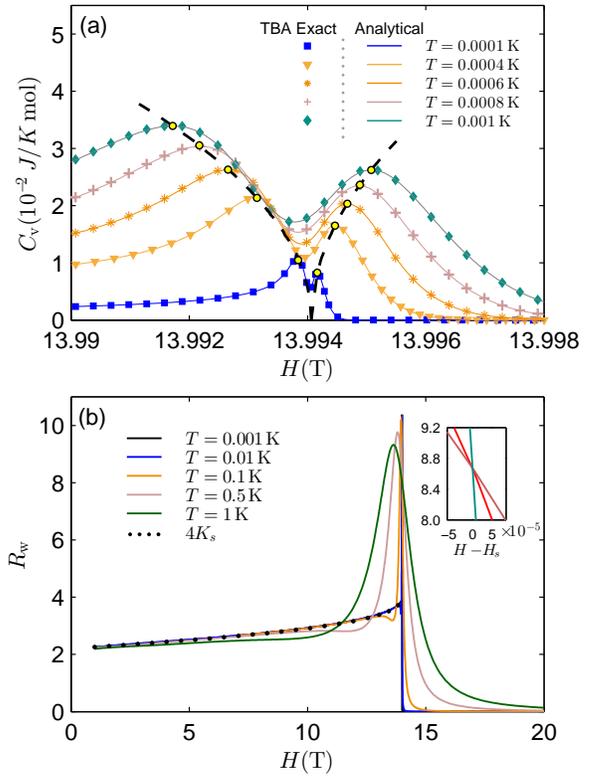}
  \end{center}
     \caption{(a) Numerical (symbols from  (\ref{TBA})) and analytical (solid lines from (\ref{free-F})) specific heat versus  magnetic field in 
     the same setting as that of 
     Fig.~\ref{Fig1}. The double-peaks (circles) fanning out from the  $Hs=13.9941$(T) mark the crossover temperatures separating the three regions: the TLL, the QC  and the FM, see Fig~\ref{Fig1}. 
   (b) A 
   numerical plot of  the Wilson ratio at different temperatures, which collapse to the Luttinger parameter curve of $4K_s$ calculated using 
   (\ref{TBA}), indicating the TLL nature. The inset shows the 
   dimensionless scaling behaviour  of the Wilson ratio  at low temperatures.  }
    \label{Fig2}
\end{figure}

{\bf Bethe ansatz equations.} The Hamiltonian of the 1D Heisenberg spin 1/2 chain  is  given by \cite{Bethe:1931}
\begin{equation}
  \mathcal{H} = 2J \sum_{j = 1}^N \vec{S}_j \cdot
  \vec{S}_{j + 1} -  g\mu_B H M^z,\label{Fig2}
\end{equation}
where 
$J$ is the intrachain coupling constant, $N$ is the number of lattice  sites and 
$M^z=\sum_{j = 1}^NS_j^z=N/2-M$ is the magnetization. $M$  is the number of down spins.  In this Hamiltonian, $g$ and  $\mu_B$ are  the Land\'{e} factor 
and the Bohr magneton, respectively.  To simplify notation, we let $g\mu_B  =1$. The spin-1/2 operator $\vec{S}_j$ associate to
the site $j$ interacts 
with its nearest neighbours under
a magnetic field  $H$.
The energy is given by
  $E= - \sum^M_{j = 1}
  \frac{J}{\lambda_j^2 + \frac{1}{4}} + H M + E_0$, 
where   $E_0 = \frac{1}{2} N \left( J - H \right)$, 
 and the spin quasimomenta 
$\lambda_j $  with $j=1,\ldots, M$ are determined by 
the Bethe ansatz (BA)  equations \cite{Bethe:1931,Takahashi:1999}, also see  \cite{Supp}.
For the ground state, all the $\lambda_j$ 
take real values. 
However, at finite temperatures and in the thermodynamic limit, 
 there are  real and complex  solutions describing different lengths of  bound states
\begin{equation}
\lambda^{n}_{j,\ell} =\lambda^n_j+\frac{1}{2}\mathrm{i}(n+1-2\ell )
\end{equation}
with $\ell =1,\ldots, n$, and $j=1, \ldots, \nu_n$. 
Here $\lambda^n_j$ and $\nu_n$ denote   the real part  and the number of 
length-$n$ strings, respectively \cite{Faddeev:1981}.

   Building on such spin strings \cite{Takahashi:1971}, the thermodynamics of the system is determined by the TBA equations
\begin{equation}
  \varepsilon_n^+=\varepsilon^0_{n}- \sum_m
  A_{m, n} \ast   \varepsilon_m^-(\lambda),\label{TBA}
\end{equation}
where $\ast  $ denotes 
convolution, $n$ takes positive integer values and 
$\varepsilon_n^\pm = \pm T \ln[1+{\rm e}^{\pm \varepsilon_n/T}]$ defines the dressed energy 
of the length-$n$ spin strings.  The driving term is given by $  \varepsilon_{n^{}}^0 = - 2 \pi J a_n(\lambda) + n H$ with the kernel function $ a_n (\lambda)= \frac{1}{2 \pi} \frac{n}{\lambda^2 + n^2 / 4}$.  The function $A_{m,n}$ is given in \cite{Supp}.
The 
free energy per unit length is given by
$  f =  \sum_n \int a_n \left( \lambda \right) \varepsilon_n^-\left( \lambda \right)d \lambda $.
Hereafter, all magnetic properties will be 
in the per unit lengths.

{\bf Spin strings and spin  liquid.}
For  low-lying excitations, each magnon  decomposes into two spinons, i.e.
spin-1/2 quasiparticles \cite{Faddeev:1981,Hammar:1999,Karbach:2000,Karbach:2002,Caux:2005,Caux:2006,Klauser:2012,YangW:2017}.
The spectral weight of two spinon excitations  have been experimentally confirmed through observation of the spin dynamic structure factor \cite{Lake:2005,Mourigal:435,Lake:2013,Zheludev:2008,Stone:2003}.
In order to calculate  the spin string contributions  to the thermodynamics at different temperature scales, we  rewrite the  free energy  as
$f =\sum_n  g_n(\lambda)
+\sum_n \varepsilon_n^-(\infty)$, 
where 
$g_n =  \int {\rm d \lambda} \: a_n(\lambda)( \varepsilon_n^-(\lambda)-\varepsilon_n^-(\infty))  $ counts the major  contribution from the length-$n$ strings, besides    their  constant values $\varepsilon_n^{\pm} (\infty)$, to 
the free energy.
Thus $g_n$  is very convenient for estimating 
the cut-off  string length $n_c$, see \cite{Supp}.
It is important to observe  that  $g_n$ shows a power law
decay as  $n$ increases,  see Fig.~\ref{Fig1}(b).

Here we observe that for  a small value of $H/T$, a large cut-off string length $n_c$ is needed  in the calculation of the thermodynamics. When $T\to \infty$, full  string patterns are required, i.e. $n_c\to \infty$, so that  the free energy reduces to that of 
free spins: $f=\sum_n \varepsilon_n^-(\infty)$. Moreover,  for $H\sim  0^+$ and $T\ll 1$,   logarithmic temperature corrections to the  thermodynamical properties of the renormalization fixed point effective  Hamiltonian have been 
seen  \cite{Johnston:2000,Lukyanov:1998,Eggert:1994}.
At $T=0$, all the $\lambda_j$ 
take real values.
In this case,  one easily gets
the known magnetization critical exponent $\delta =2$ in the scaling form   $1-M^z/M_s =D (1-H/H_s)^{1/\delta}$ with $D=4/\pi$ \cite{Supp}. This gives a divergent spin susceptibility at the saturation point $H_s=4J$ \cite{Bonner:1964}.

At low temperatures, i.e. $T\ll H$,  the  TLL  feature  is dominated by   the excitations close to the Fermi points of the length-$1$ string $\varepsilon_1$ in the parameter $\lambda$ space.  Such elementary excitations are described by particle hole excitations.
From  the TBA equations (\ref{TBA}), the dressed energy $\varepsilon_1$ is given by
$  \varepsilon_1(\lambda)=\varepsilon_1^{(0)}(\lambda)+\eta(\lambda)+O(T^3)$,
where   $\varepsilon_1^{(0)}(\lambda)$ is given by the  dressed energy equation (\ref{TBA}) in the limit $T=0$   and  the leading order  temperature  correction is determined  by
$  \eta(\lambda)=\frac{\pi^2 T^2}{6t}[a_2(\lambda+Q)+a_2(\lambda-Q)]-\int_{-Q}^{Q}a_2\ast   \eta(\mu) {\rm d} \mu$. 
 Here, $Q$ is fixed by the external field through  $\varepsilon_1^{\left( 0 \right)} \left( \pm Q \right)=0 $, see \cite{Supp}.
At low temperatures and in the limit of zero magnetic field, 
the free energy has been 
calculated by the Wiener-Hopf method \cite{Nepomechie:1993}.
For arbitrary 
$H<H_s$, we thus obtain the  field theory  result for
the free energy:
$f = E_0 - \pi^{} T^2/(6 v_s) +O(T^3)$,
where $E_0$ is the ground state energy and the sound velocity is given by $v_s = \frac{1}{2 \pi} \frac{d \varepsilon_1^{} \left( \lambda \right) / d
  \lambda}{\rho_0 \left( \lambda \right)} \mid_{\lambda = Q} $ \cite{Supp}.   This free energy  gives the relativistic behavour of phonons  \cite{Affleck:1986a},  where the specific heat is $c_v/T=\pi/(3v_s)$.
 This gives the  dynamic critical exponent $z=1$.

{\bf Quantum criticality of spinons.}
In this spin-1/2 chain, the phase transition between 
the magnetized and 
ferromagnetic phases 
occurs at the saturation  point \cite{Haldane:1981,Affleck:1986a,Maeda:2007}.
However, the determination of 
the phase boundary of the TLL  at quantum criticality  is still in question. In experiments \cite{Ruegg:2008,Kono:2015}, the magnetization peaks were 
regarded as the, as yet unjustified, TLL phase boundary. 
In  Fig.~\ref{Fig1} (a), we demonstrate that the peak positions of the specific heat( the dotted solid lines) fanning out from the saturation field $H_s$  coincide with the phase boundaries  determined by the Wilson ratio $R_{\rm W}$.   We observe  that the phase boundary of the TLL determined by the  magnetization peaks  
deviates significantly from the true TLL phase boundary as determined by the Wilson ratio and
specific heat.

In Fig.~\ref{Fig2} (a), we further demonstrate the existence of 
crossover temperatures from the double-peak structure  of the specific heat. The existence of these 
crossover temperatures results in 
three different fluctuation regions:  quantum and thermal fluctuations reach an 
equal  footing (TLL);  thermal fluctuations  strongly coupled 
to quantum fluctuations (QC);    dilute magnons dominate the fluctuations (FM). 
We show  that there exists  an intrinsic connection between the Wilson ratio and Luttinger parameter
\begin{eqnarray}
R_\mathrm{W}=4K_s\label{R-K}
\end{eqnarray}
 for  the Luttinger liquid,  i.e. $H\le H_s$, see Fig.~\ref{Fig2}(b).
Here $K_s$ is the Luttinger parameter.
A similar  relation was recently found  in spin ladder compounds and Fermi gases  \cite{Nin12,GuaYFB13,Yu:2016,Saghafi:2016}.
Thus the Wilson ratio elegantly quantifies 
the TLL regardless of the microscopic details of the underlying quantum system. 
This 
elegant relation (\ref{R-K})  is confirmed by the numerical solutions of the TBA equations (\ref{TBA}), see Fig.~\ref{Fig2} (b).
Moreover, the  relation between the Luttinger parameter $K_s$  and the sound  velocity $ K _s= \pi v_s\chi / \left( g \mu_B \right)^2$ is also universal  \cite{Giamarchi:2004}.

We further  show  that the  length-$1$ spin strings   dominate  the quantum criticality of the antiferromagnetic spin-1/2  chain in the vicinity of the critical point \cite{Supp}. We prove that the vanishing Fermi point gives rise to a universality class of free fermion criticality, i.e.\ the  dilute spinons.  By developing  the generating function of free fermions in  the TBA equations (\ref{TBA}) \cite{Supp}, we obtain the free energy
\begin{eqnarray}
  f & \approx  & - \frac{2}{\pi} b_1 + \frac{8}{\pi} b_2 \label{free-F}
  \end{eqnarray}
 near $H_s$, where
$  b_1 =  - \frac{\sqrt{\pi} T^{\frac{3}{2}}}{4 \sqrt{J}}
  \tmop{Li}_{\frac{3}{2}} \left( - e^{\frac{A}{T}} \right)$ and 
 $ b_2  =  - \frac{1}{2} \frac{\sqrt{\pi} T^{\frac{5}{2}}}{\left( 16 J
  \right)^{\frac{3}{2}}} \tmop{Li}_{\frac{5}{2}} \left( - e^{\frac{A}{T}} \right)$ with $A = 4 J - H - \frac{_{} b_1}{\pi} + \frac{b_2}{\pi}$.
This simple result gives  very accurate thermal and magnetic properties for the field  near the saturation field,  see \ref{Fig2}(a).
The polylog function $ \tmop{Li}_{3/2}(x)$ appearing in 
$b_1$ indicates that the spinons are similar in nature to free fermions. 
The magnon density $n_{\rm magnon} =M_s/N-M^z=  \frac{\sqrt{2 m^{*} T}}{\pi} \int^{\infty}_0 \frac{d x}{e^{x^2 -
  \frac{H_s - H}{T}}+1}$ 
can be obtained from (\ref{free-F}) in the vicinity of  the critical point. Here the  effective mass of the magnon is given by 
$ m^{*} \approx  \frac{1}{2J} \left( 1 - \frac{T^{1 / 2}}{ \sqrt{\pi J}}
   \int^{\infty}_0 \frac{d x}{e^{x^2 - \frac{H_s - H}{T}}+1} \right)$. 
We observe that  the effective mass decreases 
as the magnetic field moves 
away from the critical point.

\begin{figure}[t]
\vspace{0.2cm}
\begin{center}
  \includegraphics[width=0.9\linewidth]{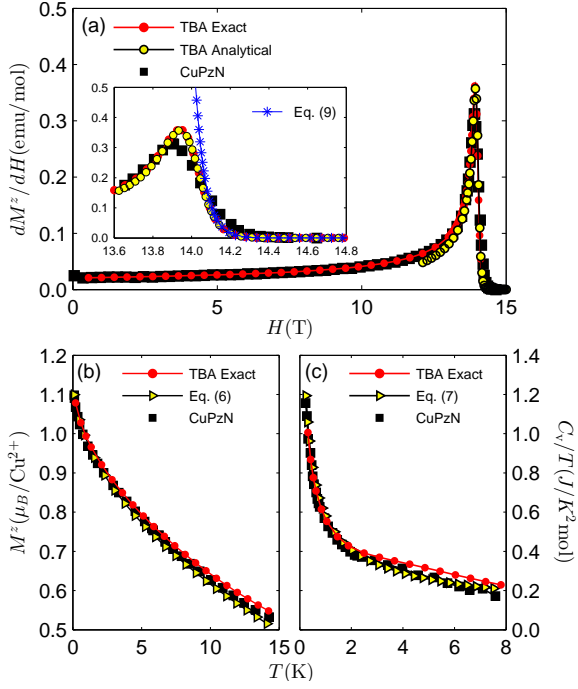}
  \end{center}
   \caption{ (a) Susceptibility versus
   magnetic field at $T=0.08$K.  The numerical (red-dots (\ref{TBA})) and analytical (yellow-circles (\ref{free-F})) results  agree well with the experimental measurement (black squares) for the 1D spin-$1/2$ antiferromagnet CuPzN \cite{Kono:2015} with the same setting used in Fig.~\ref{Fig1}.  The inset shows the exponential decay of the susceptibility, as compared  with  Eq.~(\ref{gap}), when
   the field slightly exceeds 
   the saturation field $H_s$.
  (b) and (c) show the scaling laws of the magnetization and specific heat versus 
  temperature. Excellent agreement is observed between our theoretical result  and the experimental data (black-squares), where the red-dots and  yellow-triangles denote   the numerical  TBA (\ref{TBA}) result  and the analytical scalings  Eqs.~(\ref{mag-scaling}) and (\ref{heat-scaling}), respectively.}
  \label{Fig3}
\end{figure}

Using the standard thermodynamic relations one can obtain  entire scaling  functions for the 
per unit  length magnetization and the susceptibility  for the region beyond the TLL, i.e.  $T \gg H_s-H$:
\begin{eqnarray}
M^z &=&   \frac{1}{2} + \lambda_0T^{\frac{1}{2}} f^s_{\frac{1}{2}},\qquad \chi=- \lambda_0T^{-\frac{1}{2}}f^s_{-\frac{1}{2}},
\label{mag-scaling}
\end{eqnarray}
where   $\lambda_0 =1/(2 \sqrt{\pi J})$ and $f_{n^{}}^s  =  \tmop{Li}_n \left( - e^{\frac{\Delta}{T}} \right)$ with  $\Delta=4J-H$.
These analytical scaling functions signify 
the free fermion nature of the spinons and correspond to a 
dynamical critical exponent $z=2$ and a correlation length exponent $\nu=1/2$. In particular, the magnetization
$(M_s/N-M^z)/H\propto T^{\beta }$ determines the exponent $\beta =1/2$ in the critical region.
The scaling function of the specific heat in the  critical regime is given by
\begin{eqnarray}
  c_v & = &  \sqrt{\frac{T}{\pi J}} \left[ - \frac{3}{8} f^s_{\frac{3}{2}} + \frac{1}{2}  \frac{ \Delta }{T}  f^s_{\frac{1}{2}}   - \frac{1}{2} \left(  \frac{ \Delta }{T}  \right)^2f^s_{- \frac{1}{2}}
  \right].\label{heat-scaling}
\end{eqnarray}
We see  that $ c_v/T \propto T^{- \alpha}$ with $\alpha =1/2$.
By definition, the Wilson ratio in the critical region  satisfies the scaling behaviour
$  R_W \approx   \left( \frac{4\pi k_B}{3g \mu_B} \right)^2 f_{- 1 / 2}^s/ f_{3 / 2}^s $ as $H\to H_s$.
It follows 
that  the Wilson ratio curves at low temperatures intersect, where  the slopes are proportional to $1/T$, see the inset of Fig.~\ref{Fig2} (b).

\begin{figure}[t]
\begin{center}
  \includegraphics[width=0.9\linewidth]{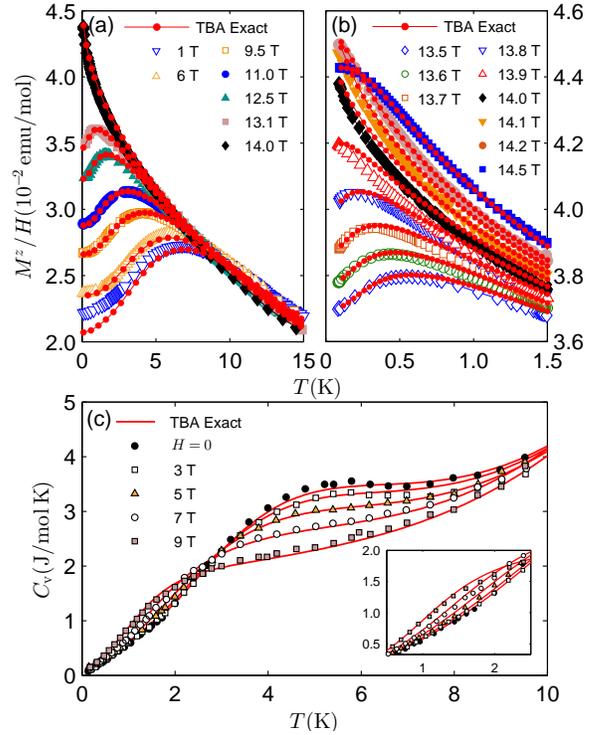}
  \end{center}
  \caption{(a)  Experimental  magnetization  $M^z/H$ versus 
  temperature at  various fields (see symbols) for the antiferromagnet CuPzN \cite{Kono:2015}. The red dots show  the TBA numerical result with the same setting used in Fig.~\ref{Fig1}. For the case $H=1.0$T,  we considered $n=120$ spin strings in order to reach a stable numerical accuracy.  (b) shows the magnetization for 
  low temperatures 
  ($T\leqslant1.5$K)  and for 
  magnetic fields near
  $H_s$, comparing 
  the numerical result (red dots) with the experimental 
  data (symbols). (c) Specific heat versus 
  temperature for CuPzN \cite{Hammar:1999} with different magnetic fields. The symbols  and solid red lines stand 
  for the experimental  and TBA 
  numerical results from (\ref{TBA}) with the cutoff string $n_c=30$. Here the phonon contribution is included. The inset shows the linear T-dependent signature within the curves as $T\to 0$.}
  \label{Fig4}
\end{figure}

So far,  we have analytically  obtained all critical exponents in the critical region:
\begin{equation}
\alpha = \beta = 1/2,\,\,\delta =2,\,\,z=2,\,\,\nu=\frac{1}{2}.
\end{equation}
They satisfy 
the relation $  \alpha + \beta \left( 1 + \delta \right) = 2 $. In addition, when 
the magnetic field slightly exceeds 
the critical field $H_s$, the ferromagnetic ordering leads to a gapped phase where the susceptibility decays exponentially, illustrating the 
 universal behaviour of the dilute magnons
\begin{equation}
\chi =  \frac{1}{2 \sqrt{\pi J T}} e^{- \Delta_g/T}\label{gap}
\end{equation}
with $ \Delta_g=4 J - H$, see Fig.~\ref{Fig3}(a).

{\bf Application to  the spin material.}
The 
analytical results obtained here for the 
quantum scaling functions (\ref{mag-scaling})--(\ref{gap})   provide a precise understanding of the quantum criticality of the ideal spin-1/2 antiferromagnet CuPzN \cite{Kono:2015}, on which 
high precision measurements 
of the thermal  magnetic properties have 
been made.
Here the best fit of magnetic properties  determines  the coupling constant $2J=10.81$K, Lande factor $g=2.3$ and the saturation field $H_s=13.9941$(T) which 
only slightly differ 
from the experimental values $2J=10.8(1)$K, $g=2.3(1)$ and $H_s=13.97(6)$(T), respectively.
Fig.~\ref{Fig3}(a) shows
excellent agreement between our theoretical results for the 
susceptibility and the experimental data for the spin-1/2 antiferromagnet CuPzN in the measured region. In particular, one can identify dilute magnon behaviour for 
magnetic fields exceeding 
$H_s$, see the inset of  Fig.~\ref{Fig3}(a). Indeed, the scaling forms of the susceptibility (\ref{mag-scaling}) and specific heat (\ref{heat-scaling}) fit quite well with the experimental data, 
see Fig.~\ref{Fig3} (b) and (c).
However, we mention a small  discrepancy between the theoretical result and experimental data for 
the susceptibility in a narrow window around 
the critical point. This is due to a 3D coupling effect, which has
also been noted in   spin ladder compounds \cite{Klanjsek:2008,Thielemann:2009,Ruegg:2008}.

In Fig.~\ref{Fig4} (a), (b), we have compared  our theoretical calculations 
with 
experimental measurements 
for 
the magnetization of 
the antiferromagnet CuPzN subjected to both weak and strong magnetic fields. 
There was  no theoretical examination on the  magnetization data  measured  in this experiment  \cite{Kono:2015}. 
Although there is 
overall agreement 
between our results 
and the data, an obvious  discrepancy between theory and experiment was 
observed for $H\sim J^{'}$ or  $H_s-H \sim J^{'}$  due to 
3D interchain coupling. For this model $J^{'}\approx 0.046$K,
see the magnetization curves at $H=14.0$, $13.9$, $13.8$T in  Fig.~\ref{Fig4} (b).
In addition, by properly choosing 
the cut-off string $n_c$, we can analyse the 
full thermodynamics of the model in the entire 
temperature regime by solving the TBA equation (\ref{TBA}). In Fig.~\ref{Fig4}(c), for the specific heat, $n_c=30$ was used.

In summary, we have analytically  obtained  scaling functions and all the critical exponents  of the thermal and magnetic properties of the spin-1/2 chain. This provides 
a rigorous theoretical understanding of the quantum criticality of spinons that has been observed in 
the antiferromagnet CuPzN \cite{Kono:2015}. We have found that the specific heat peaks elegantly mark the phase boundaries between the different phases at quantum criticality and that the Wilson ratio essentially  quantifies the TLL and characterises 
phase transition regardless of the microscopic details of the systems.
Our results also shed light on  quantum liquids and the criticality of spinons in a variety of systems of interacting bosons and fermions with internal spin degrees of freedom.

\noindent

{\em Acknowledgments.}  The authors  thank T.\ Giamarchi and H.\ Pu for helpful discussions. This work is supported by the NSFC under grant numbers 11374331 and the key NSFC grant No.\ 11534014. H.Q.L.\ acknowledges financial support from NSAF U1530401 and computational resources from the Beijing Computational Science Research Centre.

\clearpage\newpage
\setcounter{figure}{0}
\setcounter{table}{0}
\setcounter{equation}{0}
\def\thefigure{s\arabic{figure}}
\def\thetable{S.\arabic{table}}
\def\theequation{S.\arabic{equation}}
\setcounter{page}{1}
\pagestyle{plain}

\begin{widetext}
{\centering
\noindent
{\bf\Large Supplementary materials: Quantum criticality of spinons}\\
{Feng He, Yu-Zhu Jiang, Yi-Cong Yu, H.-Q.Lin, and Xi-Wen Guan}\\~\\~\\

}


{\bf I. Bethe ansatz and String hypothesis.} 

The Heisenberg spin-1/2 XXX chain is  a prototypical  integrable model, which is widely used to study quantum  magnetism in one dimension (1D). In Hans Bethe's seminal work \cite{Bethe},   a particular type of wave function,   which is called Bethe ansatz wave function,   was  proposed.  Using this  Bethe's  ansatz,  the so-called  Bethe ansatz (BA) equations and energy spectrum of the spin-1/2 XXX chain  were  given by 
\begin{gather}
	\left(\frac{\lambda _j-\frac{i}{2}}{\lambda_j + \frac{i}{2}} \right)^2=-\prod_{l=1} ^{M} \frac{\lambda_j -\lambda_l-i}{\lambda_j - \lambda_l+i}, \label{BA}\\
	E(\lambda_1,\cdots,\lambda_M)=-\sum_{j=1}^{M}\left(\frac{J}{\lambda_j ^2 +\frac{1}{4}} \right) + HM +E_0.
\end{gather}
Where $\lambda_j$ is spin quasimomentum with $j=1,\ldots,M$, and $M$ is the number of down spins.
 
The  BA equations (\ref{BA}) determine the rapidities  $\left\{\lambda_j\right\}$ which can be real and/or  complex.  The complex solutions of the Bethe roots  are called spin strings by Takahashi \cite{Takahashi:1971}
\begin{equation}
\lambda_{j,l}^n= \lambda_{j}^n +\frac{i}{2}(n+1-2l)\label{strings}
\end{equation}
with $\ell =1,\ldots, n$, and $j=1, \ldots, \nu_n$, see the main text. 
In thermodynamic limit, i.e. $N, M \to \infty$, and $M/N$ is finite, and at finite temperatures, the grant canonical description gives rise to the  so called thermodynamic Bethe ansatz (TBA) equations 
\begin{equation}
\label{TBA}
	\varepsilon_n^+ = \varepsilon_n^0 - \sum_m A_{m,n} \ast \varepsilon_n^-
\end{equation}
with $n=1,2\ldots \infty$. 
The $\ast$ here denote convolution $(a \ast b )(\lambda) = \int_{-\infty}^{\infty} a(\lambda - \mu)b(\mu) d \mu$, and $\varepsilon^{\pm}=\pm T \ln(1+e^{\pm \varepsilon_n /T})$. The driving term is $\varepsilon_n^0 =- 2\pi J a_n(\lambda) +nH=- \frac{nJ}{\lambda^2+n^2/4}+nH$ and the convolution kernel is
\begin{equation}
  A_{m, n} \left( \lambda \right) = a_{m + n} \left( \lambda \right) + 2 a_{m
	+ n - 2} \left( \lambda \right) + \cdots + 2 a_{\left| m - n \right| + 2}
\left( \lambda \right) + a_{\left| m - n \right|}.
\end{equation}
The full finite temperature  thermodynamics can be determined from the  per length free energy 
\begin{equation}
 f =  \sum_n \int_{-\infty} ^\infty a_n \left( \lambda \right) \varepsilon_n^-\left( \lambda \right)d \lambda. 
 \end{equation}

{\bf II. Magnetism at zero Temparature.} 

From the form of TBA equations (\ref{TBA}), we observe  that
$\varepsilon_n \geq 0$ for $n \geq 1$. Therefore, for $T=0$, the TBA equations and free energy per site reduce to 
\begin{eqnarray}
\label{zero_TBA}
\varepsilon_1^{\left( 0 \right)} \left( \lambda \right) & = & - 2 \pi J a_1 \left( \lambda \right) + H - \int^Q_{- Q} a_2 \left( \lambda - \mu \right) \varepsilon_1^{\left( 0 \right)} \left( \mu \right) d \mu, \\
f_0 & = & \int^{+ Q}_{- Q} a_1 \left( \mu \right) \varepsilon_1^{(0)} \left( \mu
  \right) d \mu,
\end{eqnarray}
where the $Q$ is the cut-off spin quasimomentum determined by the zero point of dressed energy, i.e.  $\varepsilon_1^{(0)} \left( \pm Q\right) = 0$. The saturation magnetic field can be easily obtained  from the condition $\varepsilon_1(0)= 0$. This gives  $H_s =4J$ and $M^z =1/2$. The zero temperature critical  properties thus can be analytically obtained for a small $Q$ near the critical field $H_s$. We can expand the zero temperature TBA equation (\ref{zero_TBA}) in terms of $\lambda$, namely 
\begin{equation}
\varepsilon_1^{(0)}(\lambda)\approx -2\pi J a_1(\lambda)+H-\frac{1}{\pi}\int_{-Q}^{Q}\varepsilon_1^{(0)}(\lambda){\rm d} \lambda \approx -2\pi J a_1(\lambda)+H - \frac{2Q(H-H_s)}{\pi}.
\end{equation}
Thus we  get $Q = \sqrt{\frac{H_s - H}{16 J}}$. The free energy, magnetization and susceptibility can directly evaluate with the zero temperature dressed energy 
\begin{eqnarray}
f_0 & \approx & \int^{+ Q}_{- Q} a_1 \left( \mu \right) \varepsilon_1^{(0)} \left(
\mu \right) d \mu \approx  \frac{4J}{\pi} \frac{(1-4Q^2)\arctan(2Q)-2Q}{1+4Q^2}+O(Q^4).
\end{eqnarray}
It follows that  the normalized magnetization and magnetic susceptibility (in per length unit)
\begin{eqnarray}
\label{Mz_zero}
M^z & =& M_s  - \frac{\partial f_0}{\partial H} = \frac{1}{2} - \frac{2}{\pi}
\left( 1 - \frac{H}{H_s} \right)^{1/2},\\
\chi &=& \frac{\partial M^z}{\partial H} = \frac{1}{2\pi} \left( \frac{1}{J
	\left( H_s - H \right)} \right)^{1/2}.
\end{eqnarray}
Using this result, we give the scaling form
\begin{equation}
1 - \frac{M^z}{M_s} = D \left( 1 - \frac{H}{H_s} \right)^{1/\delta} =
\frac{4}{\pi} \left( 1 - \frac{H}{H_s} \right)^{1 / 2}
\end{equation}
that reads off the critical exponent $\delta = 2 $ with the factor $D=4/\pi$ at zero temperature. This square-root behaviour of magnetization is showed  in Fig. \ref{Figure-SM-1}.
 
\begin{figure}[h]
	\begin{center}
	\includegraphics[width=0.4\linewidth]{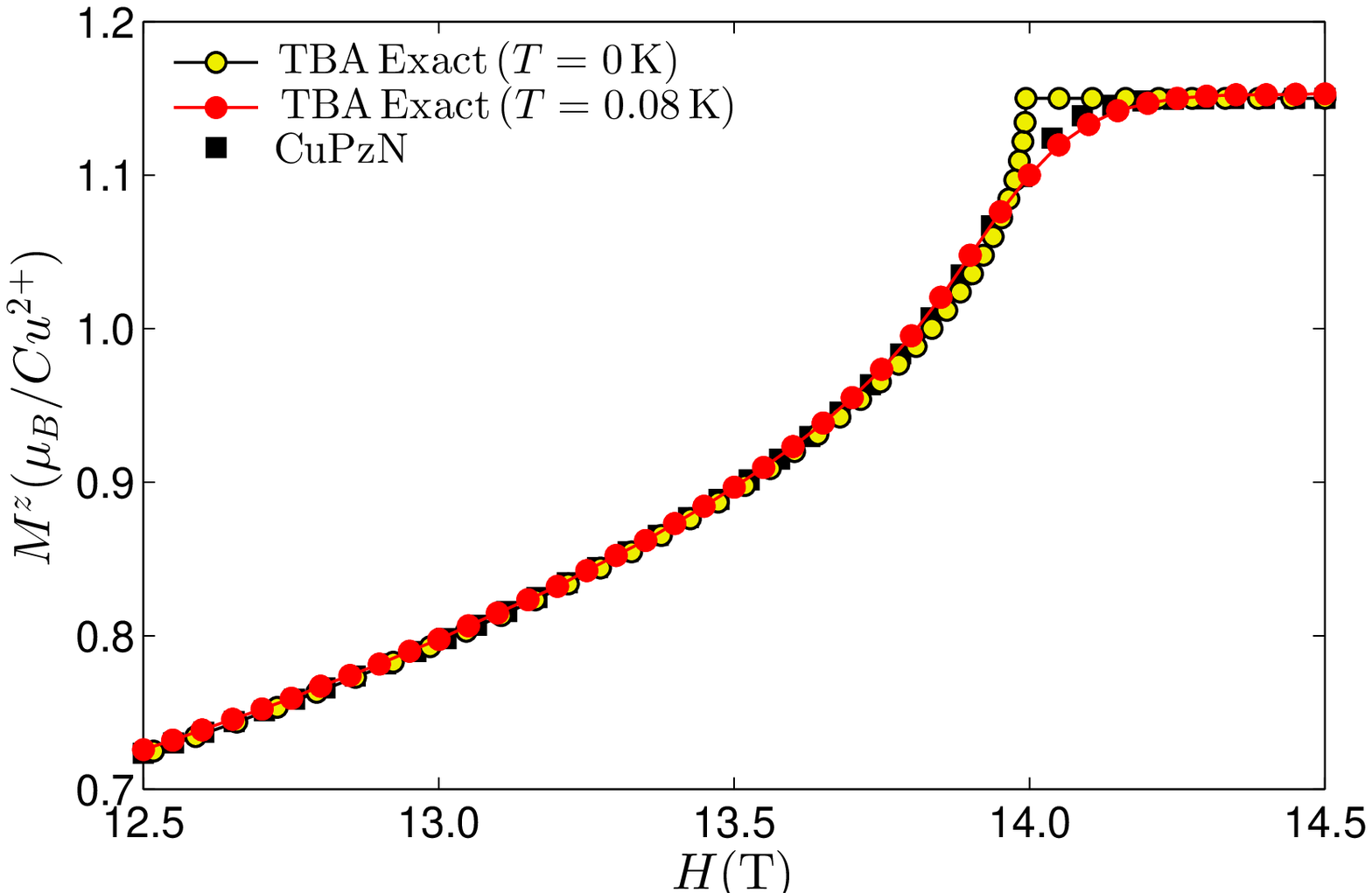}
	\end{center}
	\caption{(color online).\quad Per length magnetization $M^z$ vs external magnetic  field $H$.  The yellow-circles shows magnetization at zero temperature, which exhibits a square-root singularity at the  saturation field $H_s=13.9941(T)$. }
	\label{Figure-SM-1}
\end{figure}

{\bf III. Spin strings.} 

The low-lying excitations of the spin-1/2 system are   described by spin strings (\ref{strings}). These spin string patterns are very complicated under magnetic field and temperature.  At zero magnetic field and zero temperature, real  roots form the ground state of the spin-1/2 system. For the ground state \cite{Faddeev:1981,Caux:2005,Caux:2006,Klauser:2012} we regard the BA roots as $M=N/2$ magnons, i.e. $N/2$ length-1 spin strings  to the  BA (\ref{BA}) equations. Spin excitations are created by flipping  the dow-spins so that a magnon decomposes into two spinons carried spin-1/2.  Mathematically speaking, this spin flipping leads to two $\varepsilon_1(\lambda) $ holes in the sea of $\lambda$ roots of BA (\ref{BA}) equations.  Such a  two-spinon spectrum has been experimentally observed in many spin-1/2 systems. However, the spin excitations may lead to quite different spin string patterns, also see recent paper \cite{YangW:2017}. Here we demonstrate  three simple low-lying excitations, see Fig. \ref{Figure-SM-2}.
\begin{figure}[h]
 \begin{center}
 \includegraphics[width=0.3\linewidth]{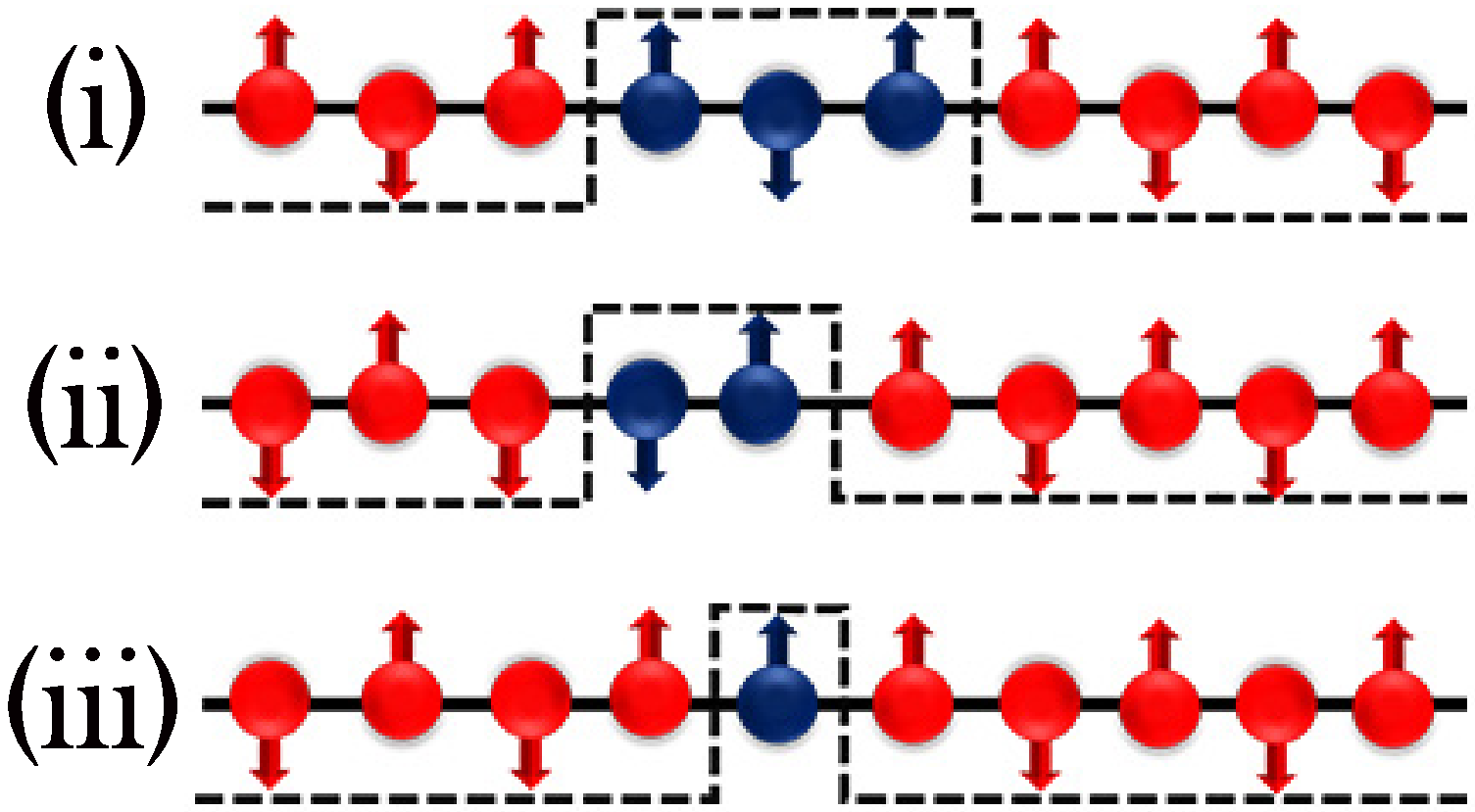}
 \includegraphics[width=0.6\linewidth]{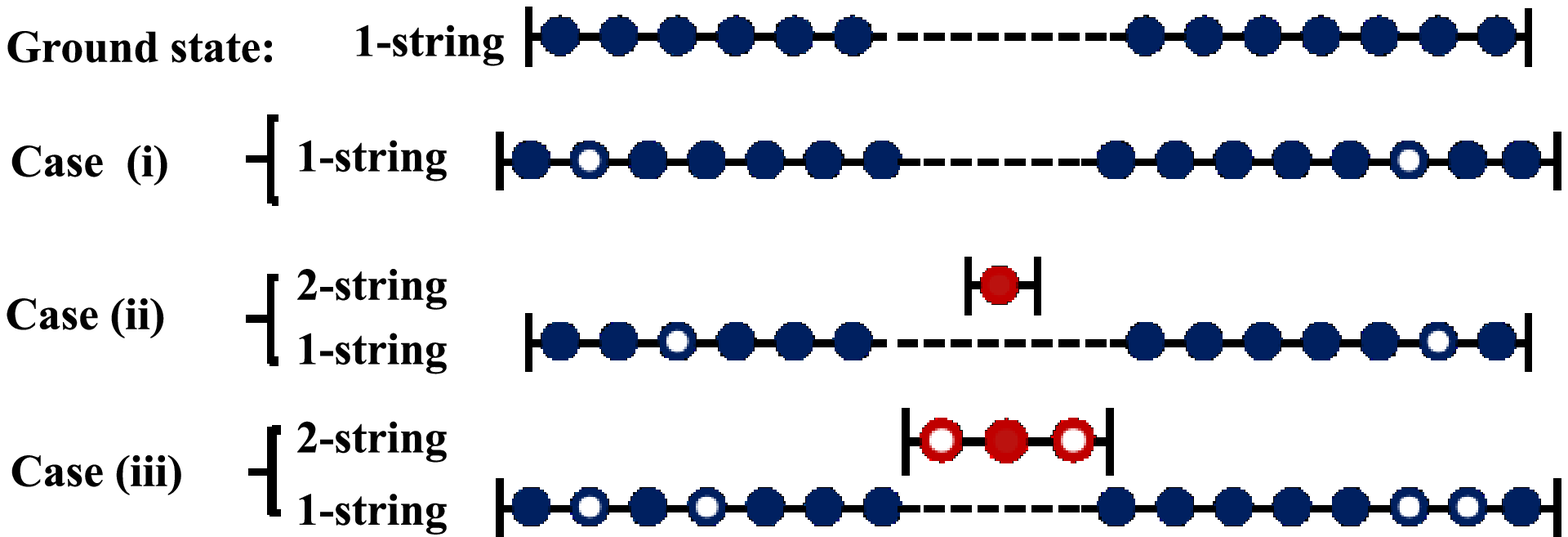}
 \caption{(color online).\quad  Schematic spin string configurations for  (i) $M^z=1$ and $2$ spinons;  (ii) $M^z=0, \nu_2 =1$ and $2$ spinons; (iii) $M^z=1, \nu_2 =1$ and $4$ spinons.}
 \label{Figure-SM-2}
 \end{center}
\end{figure}

As being  shown in Fig. \ref{Figure-SM-2}, in order to give a clear picture on the elementary excitations, we prefer to use the N\'eel state to demonstrate spin excitations over the ground state at zero magnetic field \footnote{ Note that N\'eel state is usually not  the eigenstate   of Heisenbeg antiferromagnet. Nevertheless, N\'eel state can still provide  us a  visual schematic configuration.}.

Case (i): the two-spinon excitation  with $M=N/2-1$ and  the total spin $M^z=1$. In contrast to the ground state with $N/2$ magnons, this  type of excitation has $N/2-1$  length-1 magnons  and two holes, i.e. one magnon decomposes into two magnons. Such a spin  flipping gives rise to two kinks ($\uparrow$-$\uparrow$), which are regarded as quasi-particles, i.e., two spinons. The two spinons  move with two independent rapidities.
In view of the BA equations, all vacancies are occupied for the ground state at zero magnetic filed.
One less real string  makes  the number of total vacancies increased by  one.
Therefore, in this case, the excited  states has two holes of length-$1$ string which form a scattering state of  two spinons.

Case (ii): two-spinon excitation with $M=N/2$ and total spin $M^z=0$. In this spin singlet  configuration, there is a length-$2$ string.  
Such a  singlet excitation state  is created by taking two length-$1$ strings out from the ground state pattern and add one length-2 string, see  case (ii) in Fig. \ref{Figure-SM-2}, where the two kinks ($\uparrow$-$\uparrow$ and $\downarrow$-$\downarrow$) are bounded together moving with one  velocity. The length-2 string 
 has only one vacancy.  In terms of Bethe ansatz roots, we observe that there are two spinons  in  the length-1 string sector, which define the excitation energy. This indicates that the singlet excitation also splits into two spinons.

Case (iii):  The spin triplet excitation with $M=N/2-1$ and  total spin $M^z=1$. This spin triplet excitaion is constructed by taking three length-1 strings out form the ground state pattern and add  one length-$2$  string with two holes (total three vacancies) in the length-$2 $ sector, see  case (iii) in Fig. \ref{Figure-SM-2}, where the two $\uparrow$-$\uparrow$ kinks are bounded together. The only  length-2 string occupies one of these three vacancies. These length-$2$  vacancies provide an order $\sim 1/N$ corrections  to the momentum distributions and they are negligible in thermodynamic limit. 
Based on the root patterns of  the BA equations, we observe that  there are four spinons in  the length-$1$ spin string sector. The excitation energy and momentum are  determined by these four spinons in  the thermodynamic limit.  

The above configurations  can be  obtained from the TBA equations too. 
We assume that there are $v_\nu $ length-$\nu$ strings in  the excited state. This configuration is created by taking   $\gamma$  length-$1$   strings out  of  the ground state pattern.
There is no  other length spin  strings, i.e.  $v_n=0$ for $n\neq1,\nu$.
We assume that there are $\vartheta$ holes in length-$1$ string, located at $\lambda_j ^{\rm h}$ with $j=1,2, \ldots, \vartheta $ and the density of holes in length-$1$ spin strings  is $\rho_1^{\rm h}=\frac{1}{N} \sum_{j=1}^{\vartheta} \delta (\lambda -\lambda_j ^h)$.
The $v_\nu$ length-$\nu$ strings locate at $\lambda^\nu _i$ with $i=1,2, \ldots, v_\nu $ and the corresponding density $\rho_\nu^{}=\frac{1}{N} \sum_{i=1}^{v_\nu} \delta (\lambda -\lambda^\nu_i)$.
The density of particles and holes satisfy 
\begin{equation}
 \rho_1 (\lambda) + \rho _1 ^{\rm h}(\lambda)=a_1(\lambda)- (a_2 \ast \rho_1)(\lambda) 
 -\left((a_{\nu-1}+a_{\nu+1})\ast \rho_\nu\right) (\lambda).
\end{equation}
Taking integration  with respect to $\lambda$ on both sides of  this equation, we get the number of spinons in the length-$1$ spin string sector 
\begin{equation}
 \label{ex-num}
 \vartheta =2(\gamma-v_\ell).
\end{equation}
With the help of this equation, we can find the number of holes for different kinds of spin  excitations as being discussed above.
We can also calculate the excitation energies  and momenta  by using the TBA equations.

The  spin strings configurations play important roles in  quantum dynamic process at low temperatures. However, once we consider thermodynamics of the system at   finite temperatures and finite magnetic field, contributions from different lengths of spin strings  rather  depend on numerical accuracy of the energy scales which we required.  For example, in the vicinity of the saturation point, the length-$1$ strings of magnons dominate  the critical behaviour. Different lengths of spin strings are requested to reach a certain accuracy of  energy when the magnetic field and temperature are changed.  We will further discuss the energy contributions from different  spin strings  later. 


{\bf IV. Luttinger Liquid.} 

At low temperatures, the particle-hole excitations near two Fermi points form a collective motion which is called the Luttinger liquid. Such elementary excitations only involve the roots of length-$1$ strings. Despite of differences in microscopic details between the Luttinger liquids in 1D and Fermi liquid in higher dimensions, the particle-hole excitations in 1D  lead to similar macroscopic behaviours  of higher dimensional systems at low energy.  The Luttinger liquid  behaviour can be observed in the  antiferromagnetic region with  the condition $\left| H_{} - H_s \right| / T \gg 1$. Without losing generality, we can rewrite the low temperature TBA equation (\ref{epsilon1}) as $\varepsilon_1 = \varepsilon_1^{\left( 0 \right)} + \eta$, where the $\varepsilon_1^{\left( 0 \right)}$ is zero temperature dressed energy (\ref{zero_TBA}) and $\eta$  can be regard as a leading order  correction to the temperature, namely
\begin{eqnarray}
\varepsilon_1 \left( \lambda \right) & = & - 2 \pi J a_1 \left( \lambda
\right) + H + T \int_{-\infty}^\infty a_2 \left( \lambda - \mu \right) \ln \left( 1 + e^{\frac{-
		\varepsilon_{_1} \left( \mu \right)}{T}} \right) d \mu \nonumber\\
& = & - 2 \pi J a_1 \left( \lambda \right) + H + T \left( \int^{- Q}_{-
	\infty} + \int^{\infty}_Q \right) a_2 \left( \lambda - \mu \right) \ln
\left( 1 + e^{\frac{- \varepsilon_{_1} \left( \mu \right)}{T}} \right) d \mu\nonumber\\
& & + T \int^Q_{- Q} a_2 \left( \lambda - \mu \right) \ln \left( 1 +
e^{\frac{\varepsilon_{_1} \left( \mu \right)}{T}} \right) d \mu - \int^Q_{-
	Q} a_2 \left( \lambda - \mu \right) \varepsilon_1 \left( \mu \right) d \mu
\nonumber\\
& = & - 2 \pi J a_1 \left( \lambda \right) + H + T \int^{\infty}_{- \infty}
a_2 \left( \lambda - \mu \right) \ln \left( 1 + e^{\frac{- \left|
		\varepsilon_{_1} \left( \mu \right) \right|}{T}} \right) d \mu - \int^Q_{-
	Q} a_2 \left( \lambda - \mu \right) \varepsilon_1 \left( \mu \right) d \mu. \nonumber\\
\end{eqnarray}
We then rewrite 
\begin{eqnarray}
\varepsilon_1 \left( \lambda \right) & = & \varepsilon_1^{\left( 0 \right)}
\left( \lambda \right) + \eta \left( \lambda \right) \nonumber\\
& = & - 2 \pi J a_1 \left( \lambda \right) + H - \int^Q_{- Q} a_2 \left(
\lambda - \mu \right) \varepsilon_1^{\left( 0 \right)} \left( \mu \right) d
\mu + \eta \left( \lambda \right) \nonumber\\
& = & - 2 \pi J a_1 \left( \lambda \right) + H - \int^Q_{- Q} a_2 \left(
\lambda - \mu \right) \left( \varepsilon_1 \left( \mu \right) - \eta \left(
\mu \right) \right) d \mu + \eta \left( \lambda \right).
\end{eqnarray}
It follows that 
\begin{eqnarray}
\eta \left( \lambda \right) & = & T \int^{\infty}_{- \infty} a_2 \left(
\lambda - \mu \right) \ln \left( 1 + e^{\frac{- \left| \varepsilon_{_1}
		\left( \mu \right) \right|}{T}} \right) d \mu - \int^Q_{- Q} a_2 \left(
\lambda - \mu \right) \eta \left( \mu \right) d \mu \nonumber\\
& = & I - \int^Q_{- Q} a_2 \left( \lambda - \mu \right) \eta \left( \mu
\right) d \mu
\end{eqnarray}
When $T \rightarrow 0,$ the
dominant contribution to this integration  comes from the regions near the Fermi points, i.e., the zeros of  $\varepsilon_1$. By expanding $\varepsilon_1$ at  $\lambda=Q$, we have 
\begin{equation}
\varepsilon_1 \left( \lambda \right) = t \left( \lambda - Q \right) + O
\left( \left( \lambda - Q \right)^2 \right)
\end{equation}
with $ t = \frac{d \varepsilon(\lambda)}{d \lambda} \Big |_{\lambda = Q}$. Then the  first term of $\eta$ becomes
\begin{eqnarray}
I & = & \frac{\pi^2 T^2}{6 t} \left[ a_2 \left( \lambda + Q \right) + a_2
\left( \lambda - Q \right) \right].
\end{eqnarray}
Following  a straightforward calculation, we have
\begin{equation}
\label{eta}
\eta \left( \lambda \right) = \frac{\pi^2 T^2}{6 t} \left[ a_2 \left(
\lambda + Q \right) + a_2 \left( \lambda - Q \right) \right] - \int^Q_{- Q}
a_2 \left( \lambda - \mu \right) \eta \left( \mu \right) d \mu.
\end{equation}

At zero temperature, the free energy per site $f \left( T, H \right)$ is given by 
\begin{equation}
f_0 \left( 0, H \right) = \int^Q_{- Q} a_1 \left( \lambda \right)
\varepsilon_1^{\left( 0 \right)} \left( \lambda \right) d \lambda.
\end{equation}
At low temperatures and zero magnetic field limit, the free energy was calculated by Wiener-Hopf method \cite{Nepomechie:1993}.
Here we consider  low  temperatures and finite magnetic field. Under such  conditions, the free energy is given by 
\begin{equation}
f \left( T, H \right) = - T \int^{\infty}_{- \infty} a_1 \left( \lambda
\right) \ln \left( 1 + e^{\frac{- \varepsilon_{_1} \left( \lambda \right)}{T}}.
\right) d \lambda.
\end{equation}
It follows that 
\begin{eqnarray}
f - f_0 & = & - T \int^{\infty}_{- \infty} a_1 \left( \lambda \right) \ln
\left( 1 + e^{\frac{- \varepsilon_{_1} \left( \lambda \right)}{T}} \right) d
\lambda - \int^Q_{- Q} a_1 \left( \lambda \right) \varepsilon_1^{\left( 0
	\right)} \left( \lambda \right) d \lambda \nonumber\\
& = & - T \left( \int^{- Q}_{- \infty} + \int^{\infty}_Q \right) a_1 \left(
\lambda \right) \ln \left( 1 + e^{\frac{- \varepsilon_{_1} \left( \lambda
		\right)}{T}} \right) d \lambda - T \int^Q_{- Q} a_1 \left( \lambda \right)
\ln \left( 1 + e^{\frac{\varepsilon_{_1} \left( \lambda \right)}{T}} \right)
d \lambda  \nonumber\\
& & + \int^Q_{- Q} a_1 \left( \lambda \right) \varepsilon_{_1} \left(
\lambda \right) d \lambda - \int^Q_{- Q} a_1 \left( \lambda \right)
\varepsilon_1^{\left( 0 \right)} \left( \lambda \right) d \lambda
\nonumber\\
& = & - T \int^{\infty}_{- \infty} a_1 \left( \lambda \right) \ln \left( 1
+ e^{\frac{- \left| \varepsilon_{_1} \left( \lambda \right) \right|}{T}}
\right) d \lambda + \int^Q_{- Q} a_1 \left( \lambda \right) \eta \left(
\lambda \right) d \lambda \nonumber\\
& = & - \frac{\pi^2 T^2}{3 t} a_1 \left( Q \right) + \int^Q_{- Q} a_1
\left( \lambda \right) \eta \left( \lambda \right) d \lambda
\end{eqnarray}
Then we can express  the free energy in terms  of  leading order contributions to the temperature 
\begin{equation}
\label{free_LL}
f = f_0 - \frac{\pi^2 T^2}{3 t} a_1 \left( Q \right) + \int^Q_{- Q} a_1
\left( \lambda \right) \eta \left( \lambda \right) d \lambda
\end{equation}

In order to get an close form of free energy, the key calculation is the last term in the Eq. (\ref{free_LL}). 
We use the spin-down density BA equation 
\begin{equation}
\rho_0 \left( \lambda \right) = a_1 \left( \lambda \right) - \int_{- Q}^Q
a_2 \left( \lambda - \mu \right) \rho_0 \left( \mu \right) d \mu
\end{equation}
and the Eq. (\ref{eta}), 
 we can obtain 
\begin{equation}
\int^Q_{- Q} \frac{\pi^2 T^2}{6 t} \left[ \left( a_2 \left( \lambda + Q
\right) + a_2 \left( \lambda - Q \right) \right) \right] \rho_0 \left(
\lambda \right) d \lambda = \int^Q_{- Q} a_1 \left( \lambda \right) \eta
\left( \lambda \right) d \lambda.
\end{equation}
Using the relation
\begin{eqnarray}
\rho_0 \left( Q \right) & = & a_1 \left( Q \right) - \int^Q_{- Q} a_2 \left(
Q - \mu \right) \rho_0 \left( \mu \right) d \mu, \nonumber\\
\rho_0 \left( - Q \right) & = & a_1 \left( - Q \right) - \int^Q_{- Q} a_2
\left( - Q - \mu \right) \rho_0 \left( \mu \right) d \mu,
\end{eqnarray}
and summing  up the two equations, we thus obtain 
\begin{equation}
\int^Q_{- Q}  \left[ \left( a_2 \left( \lambda + Q
\right) + a_2 \left( \lambda - Q \right) \right) \right] \rho_0 \left(
\lambda \right) d \lambda = 2 a_1 \left( Q \right) - 2 \rho_0 \left( Q
\right).
\end{equation}
Then we obtain the following result
\begin{equation}
\int^Q_{- Q} a_1 \left( \lambda \right) \eta \left( \lambda \right) d 
\lambda = \frac{\pi^2 T^2}{6 t} \left[ 2 a_1 \left( Q \right) - 2
\rho_0 \left( Q \right) \right]. 
\end{equation}
Finally, together with the formula of the free energy per site (\ref{free_LL}), we give 
\begin{eqnarray}
f & = & f_0 - \frac{\pi^2 T^2}{3 t} a_1 \left( Q \right) + \int^Q_{- Q} a_1
\left( \lambda \right) \eta \left( \lambda \right) d \lambda \nonumber\\
& = & f_0 - \frac{\pi^2 T^2}{3 t} a_1 \left( Q \right) + \frac{\pi^2 T^2}{6
	t} \left[ 2 a_1 \left( Q \right) - 2 \rho_0 \left( Q \right) \right]
\nonumber\\
& = & f_0 - \frac{\pi^2 T^2}{3 t} \rho_0 \left( Q \right) \text{}.
\end{eqnarray}
We further define sound  velocity
\begin{equation}
v_s = \frac{1}{2 \pi} \frac{d \varepsilon_1^{} \left( \lambda \right) / d
	\lambda}{\rho_0 \left( \lambda \right)} \Big |_{\lambda = Q} =
\frac{1}{2 \pi} \frac{t}{\rho_0 \left( Q \right)}.
\end{equation}
We obtain the free energy per site with  the leading order  temperature  correction 
\begin{equation}
f = f_0 - \frac{\pi^{} T^2}{6 v_s}
\end{equation}
Since $f_0$ is the free energy per site at zero temperature, it is independent of $T$. It follows that the specific heat at TLL  region is given by
\begin{equation}
c_v = - T \frac{\partial^2 f}{\partial^2 T}= \frac{\pi^{} T}{3 v_s} \propto T^{\alpha}.
\end{equation}
This  gives the exponent  $\alpha = 0$. In one dimension $\alpha = 2 - \left( d + z \right) / z$,$d=1$, so that the dynamic factor $z = 1$.


Phenomenologically, the  field theory Hamiltonian can be rewritten as   an  effective Hamiltonian in long wave length limit,  which  essentially  describes the low energy physics of the spin chain \cite{Giamarchi:2004}, namely
\begin{equation}
H = \frac{\hbar}{2 \pi} \int d x \left[ \frac{v_s K_s}{\hbar^2} \left( \pi \Pi
\left( x \right) \right)^2 + \frac{v_s}{K_s} \left( \nabla \phi \left( x
\right) \right)^2 \right],\label{H-eff}
\end{equation}
where the the canonical momenta $\Pi$ conjugate to the phase  $\phi$  obeying  the standard Bose commutation relations $\left[ \phi(x), \Pi(y) \right]=\mathrm{i} \delta(x-y)$. 
In this approach, the density variation in space  is   viewed as a superposition of harmonic waves.  
The  quantized harmonic  waves are bosons (called bosonization) and form the new eigenstate of the 1D metallic state. 
In low energy excitations, the interaction between these quantized waves are marginal.
The Luttinger parameter $K_s$ and the sound velocity $v_s$ characterize the low energy physics and determine long distance asymptotic of correlation functions. 
 Therefore the effective Hamiltonian (\ref{H-eff}) captures  the TLL physics of such kind. 

For the spin-1/2 Heisenberg chain, in the bosonization  language, the magnetization term $H_m = -g \mu_B H M^z$ in Hamiltonian can be written in term of the field $ \partial_x \phi$
\begin{equation}
H_m = \frac{ g \mu_B}{\pi} \int d x H \partial_x \phi
\end{equation}
which  is exactly the chemical potential term in the free  spinless fermions. 
Using the TLL form of the  Hamiltonian (\ref{H-eff})
the  susceptibility per length unit  is thus given by \cite{Giamarchi:2004}
\begin{equation}
\chi= \frac{- (g \mu_B)}{\pi} \frac{d \langle \nabla \phi \left( x_0
	\right) \rangle}{d H} = \frac{(g \mu_B)^2 K_s}{\pi v_s}
\end{equation}
Recalling back   the  constant factor which  we neglected, then we have 
\begin{equation}
\label{chi_Lparameter}
K_s = \frac{\pi v_s}{\left( g \mu_B \right)^2}\chi. 
\end{equation}
Whereas, for the specific heat in TLL region, we have
\begin{equation}
\label{spe_Lparameter}
c_v / T = \frac{\pi k_B^2}{3 v_s}.
\end{equation}

Moreover, the Wilson ratio are used to  characterize the interaction effect and 
spin  fluctuation. Using the relation of susceptibility (\ref{chi_Lparameter}) and specific heat (\ref{spe_Lparameter}), we obtain 
\begin{equation}
R_W = \frac{4}{3} \left( \frac{\pi k_B}{g \mu_B} \right)^2 \frac{\chi}{c_v /
	T} = \frac{4}{3} \left( \frac{\pi k_B}{g \mu_B} \right)^2
\frac{\left( g \mu_B \right)^2 K_s / \pi v_s}{\pi k_B^2 / 3 v_s} = 4 K_s.
\end{equation}
This  relation set up an intrinsic connection between the Wiilson ratio and the Luttinger parameter  for  quantum liquid. 
While this turns the phenomenological Luttinger parameter $K_s$  measurable through the Wilson ratio. 


{\bf  V. Quantum criticality. } 

For the magnetic field approaching to the saturation  filed,  the free energy and TBA equations can be  simplified  as 
\begin{eqnarray}
\label{low_TBA}
f &=& - T \int a_1 \left( \lambda \right) \ln \left( 1 + e^{\frac{-\varepsilon_{_1} \left( \lambda \right)}{T}} \right) d \lambda,  \\
\varepsilon_1 \left( \lambda \right) &=& - 2 \pi J a_1 \left( \lambda \right)
+ H + T \int a_2 \left( \lambda - \mu \right) \ln \left( 1 + e^{\frac{-\varepsilon_{_1} \left( \mu \right)}{T}} \right) d \mu. \label{epsilon1}
\end{eqnarray}
Taking an expansion with  the kernel function
\begin{equation}
a_n \left( \lambda \right) = \frac{1}{2 \pi} \frac{n}{\lambda^2 +
	n^2 /4} \approx \frac{2}{n \pi} \left( 1 - \frac{4}{n^2} \lambda^2 + \cdots \right)
\end{equation}
and after a lengthy algebra, we can obtain the free energy 
\begin{equation}
f  \approx  - \frac{2}{\pi} b_1 + \frac{8}{\pi} b_2 
\end{equation}
\begin{equation}
\varepsilon_1 \left( \lambda \right) \approx \left( 16 J - \frac{b_1}{\pi} \right)
\lambda^2 - 4 J + H + \frac{_{} b_1}{\pi} - \frac{b_2}{\pi},
\end{equation}
where we  denoted 
\begin{align}
b_1 &=  T \int \ln \left( 1 + e^{\frac{- \varepsilon_{_1} \left( \mu
		\right)}{T}} \right) d \mu,  \\
b_2 &=  T \int \mu^2 \ln \left( 1 + e^{\frac{- \varepsilon_{_1} \left( \mu
		\right)}{T}} \right) d \mu. 
\end{align}
By a  straightforward calculation with a proper  iteration via  dressed energy (\ref{epsilon1}), we find 
\begin{eqnarray}
b_1 & = & - \frac{\sqrt{\pi} T^{\frac{3}{2}}}{\left( 16 J - \frac{b_1}{\pi}
	\right)^{\frac{1}{2}}} f_{\frac{3}{2}}^{A_0}, \\
b_2 & = & - \frac{1}{2} \frac{\sqrt{\pi} T^{\frac{5}{2}}}{\left( 16 J -
	\frac{b_1}{\pi} \right)^{\frac{3}{2}}} f_{\frac{5}{2}}^{A_0}
\end{eqnarray}
with 
$A_0 = 4 J - H - \frac{_{} b_1}{\pi} + \frac{b_2}{\pi} $.
Here we  defined the function $f_n ^{A_0}= \tmop{Li}_n(-e^{\frac{A_0}{T}})$  with $ \tmop{Li}_{n} \left( x \right) =\sum_{k=1}^{\infty} \frac{x^n}{k^n}$ is the  polylogarithm function. 
Using  these expressions, we obtain the following close forms of  the dressed energy and free energy 
\begin{eqnarray}
\varepsilon_1 \left( \lambda \right) &=& \left( 16 J - \frac{b_1} {\pi} \right) \lambda^2 - 4 J + H - \frac{_{} 1}{4 \sqrt{\pi J}} \frac{T^{\frac{3}{2}}}{\left( 1 - \frac{b_1}{16 \pi J} \right)^{\frac{1}{2}}} f_{\frac{3}{2}}^{A_0}\nonumber\\
&&
+ \frac{1}{8 \sqrt{\pi J} \left( 16 J \right)} \frac{T^{\frac{5}{2}}}{\left( 1 - \frac{b_1}{16 \pi J} \right)^{\frac{3}{2}}} f_{\frac{5}{2}}^{A_0}, \\
\label{free_thermodynamics}
f & = & \frac{T^{\frac{3}{2}}}{2 \sqrt{\pi J} \left( 1 - \frac{b_1}{16 \pi J}
	\right)^{\frac{1}{2}}} f_{\frac{3}{2}}^{A_0} - \frac{T^{\frac{5}{2}}}{16 J \sqrt{\pi J} \left( 1 - \frac{b_1}{16\pi J} \right)^{\frac{3}{2}}} f_{\frac{5}{2}}^{A_0}.
\end{eqnarray}
Using standard thermodynamic relations, we can directly  calculate   magnetic  quantities, for example, the magnetization is given by 
\begin{eqnarray}
M^z &=&\frac{1}{D_m} \frac{-T^{1/2}}{2\sqrt{\pi J}} f_{1/2}^s(1-\frac{T}{8J} f_{3/2}^{s}/f_{1/2}^s) +O((T/J)^2), \label{Magnetization}\\
D_m &=&1-\frac{T^{1/2}}{\sqrt{16\pi J}}f_{1/2}^s+\frac{T^{3/2}}{2\sqrt{\pi}(16J)^{3/2}}f_{3/2}^s.
\end{eqnarray}
Here $f_n^s = \tmop{Li}_n (- {\rm e}^{\frac{4J-H}{T}})$.
In order to see free fermion nature of spinons,  we wish to express the  magnetization  (\ref{Magnetization}) as 
\begin{equation}
	M^z = M_s/N - \frac{\sqrt{2 m^{*}T}}{\pi} \int^{\infty}_0 \frac{d x}{e^{(x^2 -
			\frac{H_s - H}{T})}+1}.
\end{equation}
Here $m^{*}$ is the effective mass of the spinons. 
Using the explicit per site magnetization (\ref{Magnetization}), we can rewrite 
\begin{eqnarray*}
	M^z & \approx & M_s /N+ \frac{T^{1 / 2}}{2 \sqrt{\pi J}} \tmop{Li}_{1 / 2} \left( -
	e^{\frac{H_s - H}{T}} \right) \left[ 1 + \frac{T^{1 / 2}}{4 \sqrt{\pi J}}
	\tmop{Li}_{1 / 2} \left( - e^{\frac{H_s - H}{T}} \right) \right]\\
	& = & M_s/N - \frac{T^{1 / 2}}{\pi \sqrt{J}} \int^{\infty}_0 \frac{d
		x}{e^{(x^2 - \frac{H_s - H}{T})}+1} \left[ 1 - \frac{T^{1 / 2}}{2 \sqrt{\pi J}}
	\int^{\infty}_0 \frac{d x}{e^{(x^2 - \frac{H_s - H}{T})}+1} \right]\label{mag-T}
\end{eqnarray*}
which gives the effective mass  $ m^{*} = \frac{1}{2 J} \left( 1 - \frac{T^{1 / 2}}{\sqrt{\pi J}} \int^{\infty}_0 \frac{d x}{e^{(x^2 - \frac{H_s - H}{T})}+1} \right)$
as  $H\to H_s$. This shows the  nature of free ferimons, see a discussion \cite{Maeda:2007}.


{\bf Scaling functions.}
Near a quantum phase transition,  thermal and quantum fluctuations destroy the forward scattering process in the phase of TLL \cite{Yu:2016}. 
In the vicinity of the 
critical point $H_s$ and  $\left| H - H_s \right| / T \ll 1$, 
all magnetic properties can  be cast into universal scaling forms. 
This is called  the quantum critical region. We can obtain the scaling forms directly from  the  close form of the  free energy (\ref{free_thermodynamics}) with an extra condition $J / T \gg 1$. Then we obtain  a scaling  form of free energy in the critical region 
\begin{equation}
f \approx \frac{T^{\frac{3}{2}}}{2 \sqrt{\pi J}} \tmop{Li}_{\frac{3}{2}}
\left( - e^{\frac{4 J - H}{T}} \right).
\end{equation}
It follows that the scaling forms of the 
Magnetization and susceptibility 
\begin{eqnarray}
\label{mag_critical}
M^z &=& \frac{1}{2} + \frac{T^{\frac{1}{2}}}{2 \sqrt{\pi J}}
\tmop{Li}_{\frac{1}{2}} \left( - e^{\frac{4 J - H}{T}} \right) = \frac{1}{2}
+ T^{1 / 2} \mathcal{M}\left( \Delta H / T \right),\\
\label{sus_critical}
\chi & =& \frac{\partial M^z}{\partial H} = - \frac{1}{2 \sqrt{\pi J T}}
\tmop{Li}_{- \frac{1}{2}} \left( - e^{\frac{4 J - H}{T}} \right) = T^{- 1 /
	2} \mathcal{G} \left( \Delta H / T \right).
\end{eqnarray}
In the above equations the functions $\mathcal{M}(x) =\frac{1}{2 \sqrt{\pi J}}f^s_{1/2}(x)$,
$\mathcal{G}(x) =-\frac{1}{2 \sqrt{\pi J}}f^s_{-1/2}(x)$ are dimensionless scaling functions. 
Here we denoted 
\begin{equation}
f_{n^{}}^s  \left( \frac{\Delta }{T} \right)  =  \tmop{Li}_n \left( - e^{\frac{\Delta }{T}} \right). 
\end{equation}
where
$ \Delta = H_s - H = 4 J - H$.  
Similarly, the scaling function of the specific heat is given by 
\begin{eqnarray}
\label{spe_critical}
c_v & = & T \frac{\partial s}{\partial T} = - T \frac{\partial^2 f}{\partial
	T^2} \nonumber\\
& = & \sqrt{\frac{T}{\pi J}} \left[ - \frac{3}{8} \tmop{Li}_{\frac{3}{2}}
\left( - e^{\frac{\Delta }{T}} \right) + \frac{1}{2} \left( \frac{\Delta }{T} \right) \tmop{Li}_{\frac{1}{2}} \left( - e^{\frac{\Delta }{T}} \right)\right. \nonumber\\
	\nonumber
	&& \left.
- \frac{1}{2} \left( \frac{\Delta }{T} \right)^2 \tmop{Li}_{- \frac{1}{2}}
\left( - e^{\frac{\Delta }{T}} \right) \right] \nonumber\\
& = & T^{\frac{1}{2}} \mathcal{C} \left( \Delta H / T \right).
\end{eqnarray}
We thus read off the critical dynamic exponent  $z=2$ and correlation length exponent $\nu=\frac{1}{2}$. 
Furthermore,we can also get the scaling form of the Wilson Ratio  in critical region 
\begin{equation}
\label{Rw_critical}
R_W = \frac{4}{3} \left( \frac{\pi k_B}{g \mu_B} \right)^2 \frac{f_{- 1 /
		2}^s}{\frac{3}{4} f_{3 / 2}^s - \frac{\Delta }{T} f_{1 / 2}^s
	+ \left( \frac{\Delta }{T} \right)^2 f_{-1 / 2}^s} \approx \left( \frac{4}{3}  \frac{\pi k_B}{g \mu_B} \right)^2 \frac{f_{-1/2}^s}{f_{3/2}^s}.
\end{equation}
We compare these analytical  scaling forms of physical quantities with the numerical results calculated from the TBA equations in the Figure \ref{Figure-SM-3}.
Excellent agreement between the analytical and numerical results is seen. 

\begin{figure}[h]
	\begin{center}
	\includegraphics[width=0.8\linewidth]{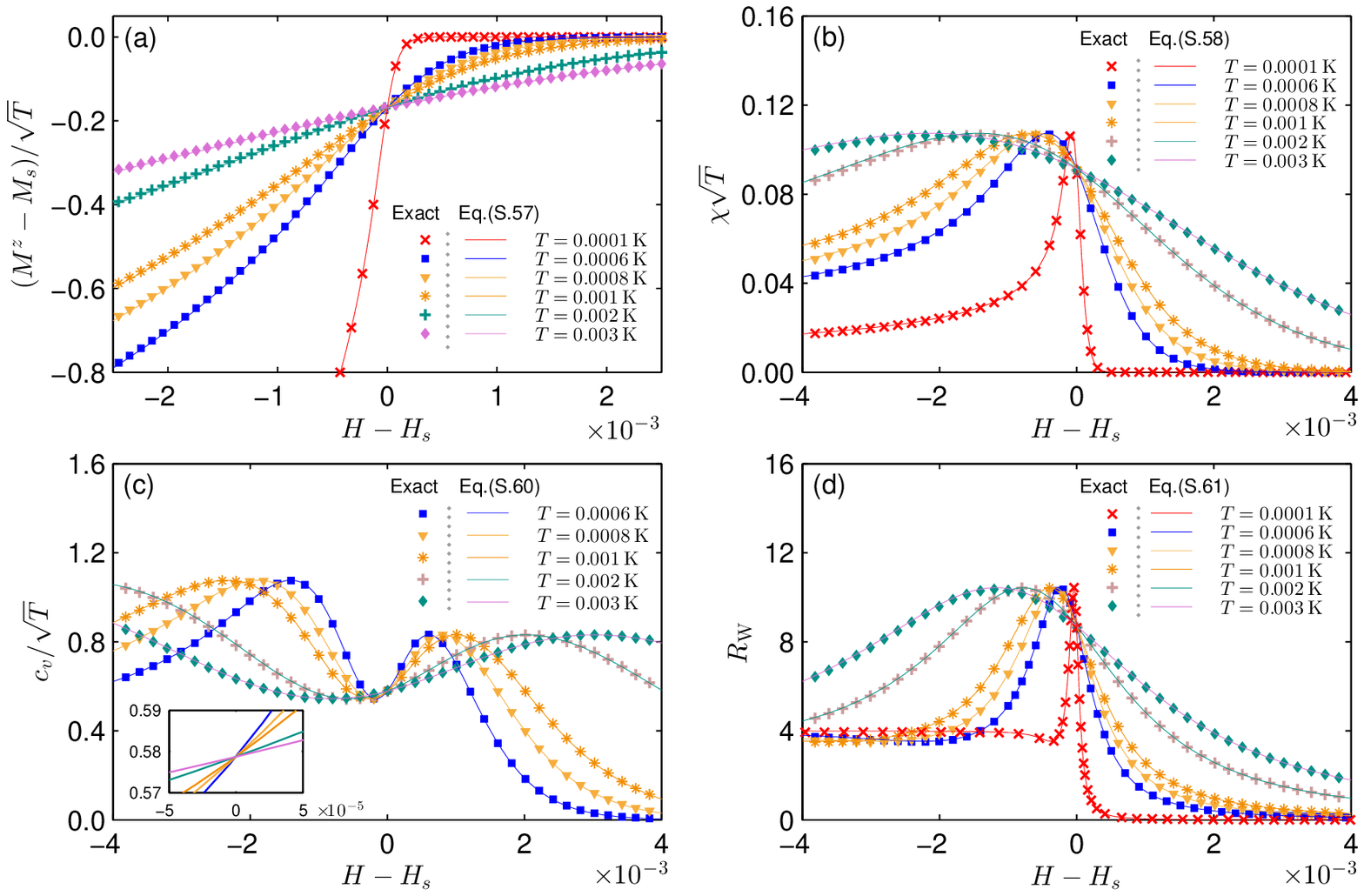}
	\end{center}
	\caption{(color online) Scaling  functions for magnetion (a), susceptibility (b), specific heat (c), Wilson ratio (d).  Analytical results  Eq. (\ref{mag_critical}), Eq. (\ref{sus_critical}),  Eq. (\ref{spe_critical}), Eq. (\ref{Rw_critical}) (lines) agree with numerical solutions of the TBA equations (\ref{TBA}). These thermodynamical  properties at different  temperatures intersect at the critical point that reads off the critical exponents, see the main text.
	 }
	\label{Figure-SM-3}
\end{figure}

{\bf Energy gap.} At zero temperature, the antiferromagnetic Heisenbeg spin chain  has a phase transition from a magnetized ground state to a ferromagnetic phase transition when  the magnetic field excess the saturation magnetic
field $H_s$. In the ferromagnetic phase an energy gap leads to  spin wave quasiparticles with a gapped dispersion. The energy gap is obtained from the TBA equations at $T\to 0$, namely 
\begin{equation}
\varepsilon_1 \left( 0 \right) = H - 4 J  = \Delta_g,
\end{equation}
where $H\ge 4J$. 
At low temperature,  the conditions $\Delta_g / T \gg 1$ always holds,  then we expand the  free energy (\ref{free_thermodynamics}), then we get 

susceptibility and  specific heat  in terms of energy gap
\begin{equation}
\chi = - \frac{1}{2 \sqrt{\pi J T}} \tmop{Li}_{- \frac{1}{2}} \left( - e^{-
	\frac{\Delta_g}{T}} \right),
\end{equation}
{\hspace{1em}}specific heat
\begin{equation}
c_v =  \sqrt{\frac{T}{\pi J}} \left[ - \frac{3}{8}\tmop{Li}_{\frac{3}{2}} \left( - e^{- \frac{\Delta_g}{T}} \right) +\frac{1}{2} \left( - \frac{\Delta_g}{T} \right) \tmop{Li}_{\frac{1}{2}}\left( - e^{- \frac{\Delta_g}{T}} \right) - \frac{1}{2} \left( -\frac{\Delta_g}{T} \right)^2 \tmop{Li}_{- \frac{1}{2}} \left( - e^{-\frac{\Delta_g}{T}} \right) \right].
\end{equation}
Taking  the limit $\lim_{\left| z \right| \rightarrow 0} \tmop{Li}_s \left( z \right) = z $, the gap equation of susceptibility and specific heat can be written as
\begin{eqnarray}
\label{chi_gap}
\chi &=& - \frac{1}{2 \sqrt{\pi J T}} \left( - e^{- \frac{\Delta_g}{T}}\right) = \frac{1}{2 \sqrt{\pi J T}} e^{- \frac{\Delta_g}{T}},\\
c_v &=& \sqrt{\frac{T}{\pi J}} \left[ \frac{3}{8} + \frac{1}{2} \left(\frac{\Delta_g}{T} \right) + \frac{1}{2} \left( \frac{\Delta_g}{T} \right)^2\right]e^{- \frac{\Delta_g}{T}}.
\end{eqnarray}
It is obviously that the susceptibility and specific show an 
exponential decay with respect to the energy gap. This nature was  directly seen  from our numerical
and experimental  fitting in the main text.


{\bf VI. Numerical solution to the TBA equations. }

The analytical expression of the dressed energy is extremely hard to derive  except for some limit cases, see the above sections.  Here we develop new  numerical method to deal with finite temperature magnetic properties of the 1D Heisenberg chain. 
The TBA equations (\ref{TBA})  consist of   infinite number of coupled integral equations of  $\varepsilon_n(\lambda)$.
In fact, it is also very difficult to solve numerically these equations.
We observe  that $\varepsilon_n(\lambda)$ approaches to a  constant for a large value of $\lambda$, i.e. 
\begin{align}
\varepsilon_n(\infty) = T \text{ln}\Big[\Big(
\frac{\text{sinh}[(n+1)H/(2T)]}
{\text{sinh}[H/(2T)]}\Big)^2-1\Big].
\end{align}
Moreover,  $|\varepsilon_n(\lambda)-\varepsilon_n(\infty)|$ decreases with increasing  the string length $n$. Thus we can take such advances to evaluate the 
quantity $\Delta \varepsilon_n^{\pm} (\lambda) = \varepsilon_n^\pm(\lambda)-\varepsilon_n^\pm(\infty)$.  In order to achieve this goal, we rewrite  the TBA equations (\ref{TBA})  as
\begin{align}
 \label{TBAmodify}
 \Delta \varepsilon_n^+(\lambda)= -2\pi J a_n(\lambda) - \sum_{m=1}^{n_{\rm c}} A_{m,n} \ast \Delta \varepsilon_n^-(\lambda) 
 - \sum_{m=n_{\rm c}+1}^{\infty} A_{m,n} \ast \Delta \varepsilon_n^-(\lambda).
\end{align}
We  choose  the cut-off string number $n_c$ large enough such that $\sum_{m=n_{\rm c}+1}^{\infty} A_{m,n} \ast \Delta \varepsilon_n^-(\lambda)$ is  negligiably  small. 
Then we are capable of performing  numerical calculation on  the dressed energies and the thermodynamic quantities. 

For the dressed energy is given by 
\footnote{
Although eq. (\ref{free-m}) can be used to calculate the free energy, it is not a good choice because of the numerical accumulation errors of $\Delta \varepsilon_n$.
The equation
\begin{align}
f= \frac{H}{2}-2J\ln2
 + \frac12\varepsilon^+_1(\infty)
 - \int {\rm d}\lambda G(\lambda) \Delta \varepsilon^+_1(\lambda)
\end{align}
gives a better numerical result.}
\begin{align}
 \label{free-m}
 &f=
 \frac{H}{2}-2J\ln 2 - T\ln[\cosh(\frac{H}{2T})]
 +\sum_{n=1}^{n_{\rm c}}g_n 
 +\sum_{n=n_{\rm c}+1}^{\infty}g_n,
 \\
 &\nonumber
 g_n = \int {\rm d \lambda} a_n(\lambda) \Delta \varepsilon_n^-(\lambda).
\end{align}
Here we find that $g_n$ decays in a power law with respect to the string length $n$
\begin{align}
 g_n|_{n\gg 1} \propto n^{-a}
\end{align} 
with  a constant exponent $a$.
For example, if we take  $k_BT/J\approx 0.2$ and $g\mu_BH/J\approx 0$,  we see $a\approx 3$.
The value of $a$ increases with respect to the  magnetic field $H$.  We observe that  $g\mu_BH/J \approx 2$, then $a\approx 10$.
This suggests  that even at the zero magnetic field limit, we still can  solve the TBA equations numerically.

In a actual  numerical process, we use $|(g_{n+1}-g_n)/g_1|<d$ to estimate the errors, where $d$ is the accuracy.
For example, we can  estimate the string length cut-off $n_c$ by setting up an  accuracy $d=10^{-6}$, see  Fig.1 in the main text.
The plateaux  feature indicates that for a certain interval of $H$, there exists a cut-off $n_c$  which  gives a high accurate numerical result with a given accuracy $d$. 
When the magnetic field $H$ is very small, higher length   strings are needed in  the numerical calculation.
For an absence of  the magnetic field, the contributions from  high length  spin strings should be taken account. In our numerical calculation, the major contributions $\sum_{n=1}^\infty \varepsilon_n^-(\infty)=\frac{H}{2}-2J\ln 2 - T\ln[\cosh(\frac{H}{2T})]$ has been already considered analytically in the above equations. We only need to calculate $\Delta \varepsilon_n^-(\lambda)$ accurately.
Upon  the accuracy $d=10^{-6}$, we find that $n_{\rm c}=11$ is enough to maintain such an accuracy. 
In particular, we would like to emphasize that near the critical point $H_s$,  we found that the  length-1 string is accurate enough to capture  the thermodynamical and magnetic  properties of the spin chain in the vicinity of the critical point $H_s$.

\end{widetext}


\begin{thebibliography}{99}


\bibitem{Yang:1966a}C. N. Yang, and C. P. Yang, Phys. Rev. {\bf 150}, 321 (1966); Phys. Rev. {\bf 150}, 327 (1966); Phys. Rev. {\bf 151}, 258 (1966).

\bibitem{Faddeev:1981}L. D. Faddeev and L. A. Takhtajan, Phys. Lett. A {\bf 85}, 375 (1981).

\bibitem{Haldane:1981}F. D. M. Haldane, Phys. Rev. Lett. {\bf 47}, 1840 (1981).

\bibitem{Affleck:1986a}A. Affleck, Phys. Rev. Lett. {\bf 56}, 2763 (1986).

\bibitem{Takahashi:1999}M. Takahashi,  {\it Thermodynamics of  One-Dimensional Solvable Models},  (Cambridge University Press,  Cambridge, 1999).


\bibitem{WangYP:2015}Y. Wang, W.-L. Yang, J. Cao, K. Shi, Off-Diagonal Bethe Ansatz for Exactly Solvable Models, (Springer-Verlag Berlin Heidelberg 2015).

\bibitem{Johnston:2000}D. C. Johnston {\em et al.}, Phys. Rev. B {\bf 61}, 9558 (2000).

\bibitem{Tennant:1995}D. A. Tennant {\em et al.}, Phys. Rev. B {\bf 52}, 13368 (1995).

\bibitem{Lake:2005}B. Lake {\em et al.},  {\bf 4}, 329 (2005).

\bibitem{Mourigal:435}M. Mourigal  {\em et al.},  Nat. Phys. {\bf 9}, {\bf 435} (2013).

\bibitem{Lake:2013}B. Lake {\em et al.}, Phys. Rev. Lett. {\bf 111}, 137205 (2013).

\bibitem{Zheludev:2008}A. Zheludev  {\em et al.}, Phys. Rev. Lett. {\bf 100}, 157204 (2008).

\bibitem{Stone:2003}M. B. Stone {\em et al.}, Phys. Rev. Lett. {91}, 037205 (2003).




\bibitem{Affleck:1986}I. Affleck, Phys. Rev. Lett. {\bf 56}, 746 (1986).

\bibitem{Cardy:1986}J. Cardy, Nucl. Phys. B {\bf 270}, 186 (1986).

\bibitem{Giamarchi:2004}T. Giamarchi {\em Quantum Physics in one  dimension} (Oxford University Press, Oxford, 2004).


\bibitem{Takahashi:1971}M. Takahashi, Prog. Theor. Phys. {\bf 46}, 401 (1971).

\bibitem{Yang:1969}C. N. Yang, and C. P. Yang,  J. Math. Phys. (N.Y.) {\bf 10}, 1115 (1969).


\bibitem{Som28}A. Sommerfeld, Z. Phys. {\bf 47}, 1 (1928).

 \bibitem{Wil75
 } K. G. Wilson, Rev. Mod. Phys. {\bf 47}, 773 (1975).



\bibitem{Kono:2015}Y. Kono {\em et al.},
  Phys. Rev. Lett. {\bf 114}, 037202 (2015).

\bibitem{Shaginyan:2016}V. R. Shaginyan {\em et al.}, Ann. Phys. (Berlin) {\bf 528}, 483 (2016).


\bibitem{Bethe:1931}H. A. Bethe,  Z. Phys. {\bf 71}, 205 (1931).

\bibitem{Supp} See supplementary material.

\bibitem{Karbach:2000}M. Karbach and G. M\"{u}ller, Phys. Rev. B {\bf 62}, 14871 (2000).

\bibitem{Karbach:2002}M. Karbach, D. Biegel and G. M\"{u}ller, Phys. Rev. B {\bf 66}, 054405 (2002).

\bibitem{Caux:2005} J.-S. Caux, R. Hagemans, and J.-M. Maillet, J. Stat.
Mech. P09003 (2005).

\bibitem{Caux:2006}J.-S. Caux and R. Hagemans, J. Stat. Mech. P12013 (2006).

\bibitem{Klauser:2012}A. Klauser, J. Mosset and J.-S. Caux,  J. Stat. Mech. P03012 (2012).


\bibitem{Lukyanov:1998}S. Lukyanov, Nucl. Phys. B {\bf 522}, 533 (1998).

\bibitem{Eggert:1994}S. Eggert, I. Affleck and M. Takahashi, Phys. Rev. Lett. {\bf 73}, 332 (1994).


\bibitem{Bonner:1964}J. C. Bonner and M. E. Fisher, Phys. Rev. {\bf 135}, A640 (1964).


\bibitem{Nepomechie:1993}L. Mezincescu and R. I. Nepomechie, {\em Quantum groups, integrable models and statistical systems}, eds. J. LeTourneux and L. Vinet, World Scientific Singapore (1993) pp 168-191;\newline
 L. Mezincescu  {\em et al.}, Nucl. Phys. B {\bf 406}, 681
(1993).

\bibitem{Maeda:2007}Y. Maeda, C. Hotta and M. Oshikawa, Phys. Rev. Lett. {\bf 99}, 057205 (2007).

\bibitem{Ruegg:2008}Ch. R\"{u}egg {\em et al.}, Phys. Rev. Lett. {\bf 101},  247202 (2008).



\bibitem{Nin12
 } K. Ninios {\em et al.}, {\bf 108}, 097201 (2012).

 \bibitem{GuaYFB13
 } X. -W. Guan {\em et al.},  Phys. Rev. Lett. {\bf 111}, 130401 (2013).

 \bibitem{Yu:2016}Y.-C. Yu  and Y.-C. Chen,  H.-Q. Lin,   R. A. Roemer, and X.-W.  Guan,  Phys. Rev. B {\bf 94}, 195129 (2016).

\bibitem{Saghafi:2016}Z. Saghafi {\em et al.}, J. Mag. Mag. Mat. {\bf 398}, 183 (2016).



\bibitem{Klanjsek:2008}M. Klanjsek {\em et al.}, Phys. Rev. Lett. {\bf 101},  137207 (2008).

\bibitem{Thielemann:2009}B. Thielemann {\em et al.}, Phys. Rev. Lett. {\bf 102},  107204 (2009).



\bibitem{Hammar:1999} P. R. Hammar {\em et al.}, Phys. Rev. B {\bf 59}, 1008 (1999).




\end{thebibliography}

\begin{thebibliography}{99}

\bibitem{Bethe} H. Bethe, {\em Z. Physik} {\bf 71}, 205 (1931).
\bibitem{Takahashi:1971}M. Takahashi, Prog. Theor. Phys. {\bf 46}, 401 (1971).
\bibitem{Faddeev:1981}L. D. Faddeev and L. A. Takhtajan, Phys. Lett. A {\bf 85}, 375 (1981). 

\bibitem{Caux:2005} J.-S. Caux, R. Hagemans, and J.-M. Maillet, J. Stat.
Mech. P09003 (2005).

\bibitem{Caux:2006}J.-S. Caux and R. Hagemans, J. Stat. Mech. P12013 (2006).

\bibitem{Klauser:2012}A. Klauser, J. Mosset and J.-S. Caux,  J. Stat. Mech. P03012 (2012). 

\bibitem{YangW:2017}W. Yang, J. Wu, S. Xu, Z. Wang and C.-J. Wu, arXiv:1702.01854. 

\bibitem{Nepomechie:1993}L. Mezincescu and R. I. Nepomechie, {\em Quantum groups, integrable models and statistical systems}, eds. J. LeTourneux and L. Vinet, World Scientific Singapore (1993) pp 168-191;\newline
 L. Mezincescu  {\em et al.}, Nucl. Phys. B {\bf 406}, 681
(1993).

\bibitem{Maeda:2007}Y. Maeda, C. Hotta and M. Oshikawa, Phys. Rev. Lett. {\bf 99}, 057205 (2007).

\bibitem{Giamarchi:2004}T. Giamarchi, {\em Quantum Physics in one  dimension} (Oxford University Press, Oxford, 2004).

 \bibitem{Yu:2016}Y.-C. Yu  and Y.-C. Chen,  H.-Q. Lin,   R. A. Roemer, and X.-W.  Guan,  Phys. Rev. B {\bf 94}, 195129 (2016). 

\end{thebibliography}
\end{document}